\documentclass[preprint,prb,aps]{revtex4}
\usepackage{amsmath,amssymb}
\usepackage{graphicx}

\begin{document}


\title{Molecular Dynamics Simulation of Solvent-Polymer Interdiffusion. I. 
Fickian diffusion}
\author{Mesfin Tsige\footnote{mtsige@sandia.gov}}
\author{Gary S. Grest\footnote{gsgrest@sandia.gov}}
\affiliation{Sandia National Laboratories, Albuquerque, NM 87185}
\date{\today}


\begin{abstract}
The interdiffusion of a solvent into a polymer melt has been studied using
large scale molecular dynamics and Monte Carlo simulation 
techniques. The solvent concentration profile and weight gain by the polymer
have been measured as a function of time. The weight gain is found to scale as
$t^{1/2}$, which is expected for Fickian type of diffusion. The concentration
profiles are fit very well assuming Fick's second law with a constant
diffusivity. The diffusivity found from fitting Fick's second law is found to be
independent of time and equal to the self diffusion constant in the dilute
solvent limit. We separately calculated the diffusivity as a function of concentration
using the Darken equation and found that the diffusivity is essentially constant
for the concentration range relevant for interdiffusion.
\end{abstract}


\maketitle

\section{Introduction}
%
The interdiffusion of a solvent into a polymer has been a subject of 
experimental and theoretical research due to both its scientific and practical 
importance. There has been a number of studies of penetrant
diffusion in homopolymers\cite{vrentas,sonnenburg,takeuchi,pant,muller92,sok,gusev,tamai,gee,hofman,li,muller98,griffiths,hahn,vegt,greenfield,tocci,lim} but less on interdiffusion of solvent into 
polymer\cite{durning,hassan,sanopoulous,stamatialis,mesfin}
or polymer-polymer interdiffusion.\cite{jilge,deutsch} 
Predicting accurately the nature of the interdiffusion of a solvent into a 
polymer film has turned out to be a challenging problem due to the large number
of factors that control the diffusion process, including the molecular weight 
distribution of the polymer and size of the solvent.\cite{pant}
Whether the polymer is a melt above its glass transition or an amorphous solid
below the glass transition significantly changes the interdiffusion process.

\begin{figure}[bth!]
\begin{center}
\includegraphics*[width=2.0in]{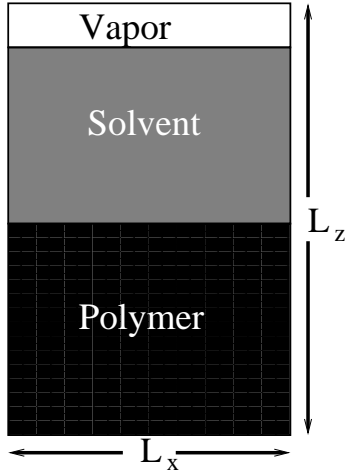}
\end{center}
 \caption{
Schematic representation of the MD simulation box for the interdiffusion study.
To allow diffusion in one direction the system is periodic in the 
$x$ and $y$ direction but not in $z$. The vapor region is added to keep the pressure 
constant.
}
\label{fig:setup}
\end{figure}

For a polymer melt, if a solvent film is placed in contact with the polymer
as shown in Fig.~\ref{fig:setup}, then the
diffusion is one-dimensional and can often be described satisfactorily
by Fick's second law\cite{crank}
\begin{equation}
\frac{\partial c}{\partial t}=\frac{\partial}{\partial z}(D(c)\frac{\partial c}{\partial z}),
\label{fick}
\end{equation}
where $c$ is the solvent concentration and $D(c )$ is the diffusivity. This 
equation assumes that the volume of the medium is not changed by the 
interdiffusion of the solvent. In this case the nature of the diffusion is 
called Fickian or Case I.\cite{kuipers,thomas2} One fingerprint for Fickian 
diffusion is that the penetration or weight gain by the polymer system 
increases as $t^{1/2}$. In general $D(c)$ is dependent of concentration 
$c$ and Eq.~\ref{fick} must be solved numerically, except in special 
cases.\cite{crank} If the diffusivity $D(c)=D_o$ is independent of
solvent concentration $c$ then the solution of Eq.~\ref{fick} for the 
concentration of solvent in the medium as a function of time and position is 
simply
\begin{equation}
c(z,t)=c_0\left(1-\text{erf}\left(z/2\sqrt(D_ot)\right)\right).
\label{concentration}
\end{equation}
Here $c_0$ is the equilibrium solvent concentration in the polymer usually
expressed in units of mass per unit volume and erf is 
the error function. Even though the 
solvent in general swells the polymer and $D(c)$ may not be independent of $c$, 
the simple functional form Eq.~(2) is often used to fit experimental data.\cite{isabel}

For glassy polymers the diffusion process does not always follow 
the standard Fickian model and is, in general, referred to be anomalous or
non-Fickian diffusion phenomena, which is caused by viscoelastic effects 
in the polymer-solvent system. One type of anomalous behavior, which has been
observed experimentally, is called 
Case II\cite{thomas2,thomas1,thomas3,hui,stamatialis} in which the polymer
relaxation process is very slow compared to the diffusion and exhibits a 
sharp concentration front that propagates at constant speed. However, recent
careful experiments\cite{stamatialis,sanopoulous,durning,hassan} have shown that 
a Fickian-like precursor foot 
proceeds the sharp front. To our knowledge no simulations have been done to date
on Case II
diffusion due to the extensive computational effort required. In this paper we
report on our interdiffusion studies of a solvent into a homopolymer 
above the glass transition 
temperature, which is expected to exhibit Fickian diffusion behavior. We leave the 
anomalous interdiffusion into a glassy polymer for a later study.

The transport of penetrant molecules in rubbery polymers has been studied using
molecular dynamics (MD) simulation techniques.\cite{sonnenburg,takeuchi,pant,
muller92,sok,tamai,gee,hofman,li,muller98,hahn,vegt,lim}
Most of these studies have focused on the penetrant diffusion of solvent
molecules in a polymer to determine the self diffusion constant.
The self diffusion constant $D_s(c)$ of the solvent is easily
calculated from the slope of the solvent mean square displacement
for long times according to the Einstein relation
\begin{equation}
D_s(c)=\lim_{t \rightarrow \infty}\frac{\langle[\boldsymbol{r}(t)-\boldsymbol{r}(0)]^2\rangle}{6t}.
\label{tracer}
\end{equation}
In eqn.~\ref{tracer} the $\langle...\rangle$ denote an ensemble average
and is obtained by averaging over all solvents and many initial time origins.
In general $D(c)$ and $D_s(c)$ are equal only in the dilute solvent limit, 
$c\simeq0$. 

There have been a number of proposed relations between $D(c)$ and $D_s(c)$, the
most common is the Darken equation\cite{darken}
\begin{equation}
D(c)=D_c(c)\left(\frac{\partial\ln f}{\partial\ln c}\right)_T
\label{darken}
\end{equation}
where $D_c(c)$ is called the corrected diffusivity and is related to molecular
mobility and $f$ is the fugacity of the solvent in the polymer. The 
thermodynamic factor $\partial \ln f/\partial \ln c$ goes to unity in the
dilute limit. Eq.~\ref{darken}
assumes that diffusion is driven by gradients in chemical potential. The 
corrected diffusivity can be expressed microscopically as\cite{sharma,maginn,neogi}
\begin{equation}
D_c(c)=\frac{1}{3Nx_sx_p}\int_{0}^{\infty}\langle \boldsymbol{J}(t)\cdot \
\boldsymbol{J}(0)\rangle dt,
\label{corrected}
\end{equation}
where $\boldsymbol{J}(t)$ is the interdiffusion current, 
\begin{equation}
\boldsymbol{J}(t)=x_p\sum_{i=1}^{N_s}\boldsymbol{v}_i(t)-x_s\sum_{j=1}^{N_p} \
\boldsymbol{v}_j(t).
\label{current}
\end{equation}
Here  $x_s$ and $x_p$ are the mole fractions of the solvent and the polymer, 
respectively and $N_s$ and $N_p$ are the number of solvent and polymer monomers.
The Einstein form of Eq.~\ref{current} is\cite{neogi}
\begin{equation}
D_c(c)=Nx_sx_p\lim_{t\rightarrow \infty}\frac{1}{6t}\langle \{\left[\boldsymbol{r}_{cm,s}(t)-\boldsymbol{r}_{cm,s}(0)\right]-\left[\boldsymbol{r}_{cm,p}(t)-\boldsymbol{r}_{cm,p}(0)\right]\}^2\rangle
\label{einstein}
\end{equation}
where $\boldsymbol{r}_{cm,s}(t)$ and $\boldsymbol{r}_{cm,p}(t)$ are the center
of mass of all solvent monomers and all polymer molecules at time $t$, respectively. The Einstein form is preferable to the Green-Kubo form (velocity auto correlation) since it avoids the need to integrate the time correlation functions.

To the best of our knowledge, no MD simulation
has been carried out to study the relationship between the interdiffusion
of a solvent into a polymer and the diffusivity and self diffusion of solvent in an
equilibrated polymer solution. Unlike solvent diffusing in a zeolite\cite{maginn,skoulidas,chandross} or
other microporous media,\cite{macelroy} the polymer swells as the solvent interpenetrates,
making it difficult to measure the concentration dependence of the
diffusivity $D(c)$ directly. 

Here we are mainly interested in investigating the penetration rates and 
concentration profiles as a function of time through direct analysis of 
molecular trajectories for Fickian diffusion in amorphous polymer-solvent 
systems. We are particularly interested in the relation between the diffusivity
$D(c)$, the self diffusion constant $D_s(c)$ and the corrected diffusion 
constant $D_c(c)$ for the solvent.
Under these circumstances, MD is a useful tool to determine the desired quantities and we have used it for our present study. By combining MD simulation with
Monte Carlo\cite{aidan} methods we can determine all of these quantities.

The paper is organized as follows. In the following section a brief review of 
the model and the simulation method used is given. The two types of thermostats,
Langevin and Dissipative Particle Dynamics (DPD), used in the simulation are also
briefly described. In Sec. III the results for the diffusion constants as a
function of solvent concentration
using the two thermostats are presented. In Sec. IV the interdiffusion results
are presented and discussed. Finally, the main results are summarized in Sec. V.

\section{Model and Simulation Details}

We performed MD simulations of polymer-solvent system using the coarse grained
bead-spring model which has been applied successfully to study the effect of
entanglement in polymer melts.\cite{kremer1} In this model the polymers are represented by freely jointed bead spring chains of length $N$ monomers of
mass $m$. The solvent is modeled as either single monomers of mass $m$ or dimers
of mass 2$m$.
The potential energy associated with interaction between nonbonded monomers of type $\alpha$ and $\beta$ is given by the standard Lennard-Jones 6-12 potential
\begin{equation}
U_{LJ}(r)=\left\{ \begin{array}{ll}
4\epsilon_{\alpha\beta}\left\{\left(\frac{\sigma_{\alpha\beta}}{r}\right)^{12}-\left(\frac{\sigma_{\alpha\beta}}{r}\right)^{6}\right\}+\epsilon_{LJ}, & r\le r_c\\
0, & r>r_c
\end{array} \right.
\end{equation}
where $r$ is the distance between monomers. Here we take $\sigma=\sigma_{\alpha\beta}$ and $r_c=2.5\sigma$. $\epsilon_{\alpha\beta}$ defines the units of energy. In this study we set $\epsilon=\epsilon_{pp}=\epsilon_{ss}$ where $p$ stands for a polymer monomer
and $s$ for a solvent monomer and vary the relative interaction $\epsilon_{sp}$.

In addition to the Lennard-Jones interaction between bonded monomers we add an anharmonic interaction term known as FENE potential,
\begin{equation}
U(r)=\left\{ \begin{array}{ll}
-0.5R_0^2k\ \text{ln}\left[1-\left(r/R_0\right)^2\right], & r\le R_0\\
\infty, & r<R_0
\end{array} \right.
\end{equation}
where, as in previous studies\cite{kremer1,grest} $R_0=1.5\sigma$ and $k=30\epsilon$. 

Our system is also weakly coupled to a heat bath in order to keep it at the desired temperature
and as a result each monomer moves according to the following stochastic equation of motion
\begin{equation}
m\frac{d^2\boldsymbol{r}_i}{dt^2}=-\boldsymbol{\nabla} \sum_{j\neq i}U(r_{ij})+\boldsymbol {F}_i^D+\boldsymbol{F}_i^R,
\label{motion}
\end{equation}
where $m$ is the monomer mass, $U(r_{ij})$ is the sum of the Lennard-Jones and an
harmonic spring potential and $\boldsymbol{F}_i^D$ and $\boldsymbol{F}_i^R$ are the dissipative force and
random force, respectively. These later two terms, $\boldsymbol{F}_i^D$ and $\boldsymbol{F}_i^R$ together
define the type of thermostat used in the simulation. In many simulations of 
polymer melts, a Langevin thermostat is used in which the particles are coupled
weakly to a heat bath. For polymer melts this is a good way to thermostat the 
system since long range hydrodynamic interactions are screened.
However these same interactions are important in the diffusion of small 
molecules as in the pure solvent. For this reason we have used two types of 
thermostats: the Langevin thermostat, which by its nature screens the 
hydrodynamics interactions and the Dissipative Particle Dynamics (DPD) 
thermostat, which does not.\cite{dunweg,soddemann} Part of the aim of this study is to compare
the results for the two thermostats.

\subsection{Langevin Thermostat}

In the Langevin thermostat the dissipative part of the force takes the form of friction that is proportional to the monomer velocity
\begin{equation}
\boldsymbol{F}_i^D=-m\gamma\frac{d\boldsymbol{r}_i}{dt},
\end{equation}
where $\gamma$ is the damping constant which is the same for all monomers. The random force is
related to the frictional force by the fluctuation/dissipation theorem
\begin{equation}
\langle \boldsymbol{F}^R_i(t)\cdot \boldsymbol{F}^R_j(t') \rangle =6\gamma k_B Tm\delta_{ij}\delta(t-t'); \langle \boldsymbol{F}^R_i(t) \rangle =0,
\end{equation}
where $k_B$ is Boltzmann's constant and $T$ is the temperature. The damping constant $\gamma$
controls both the variance of the random force and the magnitude of the frictional force and 
ensures that the system is kept stable at the desired temperature through out the simulation. 
However, since the frictional force does not conserve momentum, the hydrodynamic interactions are screened.

\subsection{Dissipative Particle Dynamics (DPD) Thermostat}

DPD has been first introduced\cite{hoogerbrugge} and applied to different systems\cite{espanol,groot,gibson,gibson2,malfreyt,elliott,ripoll} as a mesoscopic simulation
technique (not only as a thermostat) to simulate hydrodynamic behavior as well as the rheological properties of complex fluids when thermal fluctuations are important. The fluid is modeled in terms of mesoscopic particles known as dissipative particles that are large quasi-particles, which evolve in the same way as MD particles do, but with different inter-particle
interactions that allow for much longer time steps. The forces between each pair of dissipative particles is made up of a conservative force, a dissipative force and a random force similar to the one given by Eq.~\ref{motion}, but each of which is pairwise additive. These forces in general
conserve total momentum and have a spatial range given by the cut-off distance $r_c'$. In this case, the
dissipative and random forces are the main ingredients for hydrodynamic interaction and are commonly given by,
\begin{equation}
\boldsymbol{F}_{ij}^D=-m\gamma \omega^D(r_{ij})(\hat{\boldsymbol{r}}_{ij}\cdot \boldsymbol{v}_{ij})\hat{\boldsymbol{r}}_{ij}, \qquad 
\boldsymbol{F}_{ij}^R=m\sigma\omega^R(r_{ij})\zeta_{ij}\hat{\boldsymbol{r}}_{ij},
\end{equation}
where $\gamma$ and $\sigma$ are constants, $\omega^D(r)$ and $\omega^R(r)$ are weight functions that vanishes for $r > r_c'$, 
$\boldsymbol{v}_{ij}$ and $\boldsymbol{r}_{ij}$ are relative velocity and distance between the pairs, respectively,
$\zeta_{ij}$ is a random noise term with zero mean. The random number can be sampled from either
a Gaussian or a uniform distribution, in each case with a variance of unity. Espa\~nol
and Waven\cite{espanol} have showed that although the weight function can be 
chosen arbitrarily, it must satisfy the relation
$\omega^D(r)=[\omega^R(r)]^2$.
The friction coefficient $\gamma$ and the noise amplitude $\sigma$ are related 
by the fluctuation/dissipation theorem, $m\sigma^2=2\gamma k_BT$.

In the present study, however, we used DPD to thermostat the system to take advantage of the fact that it conserves the 
hydrodynamic interaction between the particles. The function used for $\omega^D(r)$ and 
$\omega^R(r)$ in this study have the following form\cite{elliott}
\begin{equation}
\omega^D(r)=[\omega^R(r)]^2=\left\{ \begin{array}{ll}
\left(1-r/r_c\right)^2, & r < r_c'\\
0, &  r \ge r_c'
\end{array} \right.
\label{omega}
\end{equation}
where we used the same cutoff as for the 
Lennard-Jones interaction cutoff between monomers, $r_c'=r_c=2.5\sigma$.

All the simulations were run using the massively parallel code 
LAMMPS.\cite{steve} The equations of motion were integrated with a 
velocity-Verlet algorithm with a time step $\Delta t=0.012\tau$ for the
interdiffusion study and  $\Delta t=0.009\tau$ for the bulk equilibrium
measurement of the self and corrected diffusion constants, where
$\tau =m(\sigma/\epsilon)^{1/2}$. The smaller time
step was used to assure that the system was stable even for large $\gamma$ (see
Fig.~\ref{fig:tracer}). All the simulations were carried out at
a temperature of $T=\epsilon/k_B$ and pressure $P\simeq0$. To determine the 
fugacity f, the LADERA grand canonical molecular dynamics (GCMD) code was 
used.\cite{aidan} During the course of an equilibrium
molecular dynamics simulation at the appropriate solvent concentration,
the energy of inserting a solvent particle at random locations was sampled.

The simulated system for the interdiffusion study consists of a rectangular
cell, which is periodic in $x$ and $y$ direction but not in $z$ as shown
in Fig.~\ref{fig:setup}. This initial configuration was generated in two 
steps. First, a polymer melt system ($L_x=L_y=60\sigma$ and thickness 
$L_z\simeq96\sigma$) was equilibrated between two walls at pressure $P\simeq0$.
Then the top of the box was extended and the solvent molecules
were placed in contact with the polymer melt. Furthermore to keep the pressure
of the system constant, a small vapor phase is added between the top wall 
and the solvent particles. The polymer melts in this study consisted of 
chains of length $N=500$ or $50$ monomers. The total number of polymers 
monomers in all cases was 300,000. The solvent consisted of either 
230,000 monomers or 193,000 dimers.

For studying self and corrected diffusion as a function of solvent 
concentration, the system consists of an equilibrated polymer solvent
mixture in a cubic cell, which is periodic in all three directions. The total
number of polymer monomers $N_p=50,000$ in 100 chains each with 500 monomers. 
The number of solvent monomers $N_s$ in the system was varied from 500 monomers
(dilute case) to 150,000 monomers. A pure solvent system of 50,000 monomers
was also simulated. As the concentration of solvent is varied, the
pressure in the system is kept constant by allowing the system relax to the
corresponding volume, thus fixing the volume for the measurement of the 
diffusion constant. Only a few hundred thousand MD timesteps are required to 
calculate the self diffusion constant $D_s(c)$ though more than ten times this 
is required for the corrected diffusion constant $D_c(c)$. The
fugacity calculation requires approximately a million MC cycles and a few 
hundred thousand MD timesteps especially at low solvent concentration.

\section{Diffusion Coefficients}

\subsection{Dependence of self diffusion on the strength of the dissipative
force $\gamma$} 
\begin{figure}[bth!]
\begin{center}
\includegraphics*[width=3.0in]{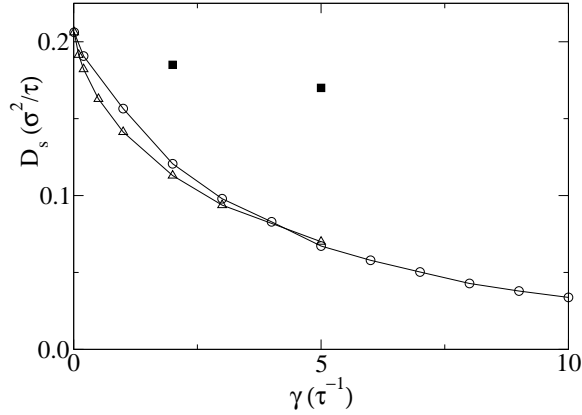}
\end{center}
 \caption{
Dependence of self diffusion constant $D_s(1)$ for the pure solvent on 
the damping coefficient $\gamma$ for the 
Langevin ($\vartriangle$) and DPD with $r_c'=2.5\sigma$ (o) and $r_c'=2.0\sigma$ ($\blacksquare$) thermostats for $T=\epsilon/k_B$ and 
$P\approx0$. The solid lines are a guide to the eye.  Error bars are
$\pm0.002\sigma^2/\tau$.
 }
 \label{fig:tracer}
 \end{figure}

A set of simulations were performed to determine the dependence
of the diffusion on $\gamma$. The self diffusion constant, $D_s(c)$, of the solvent in the
system is calculated using Eq.~\ref{tracer}.  The
results for $D_s(1)$ are shown in Fig.~\ref{fig:tracer} for the two thermostats.

We find that the values of $D_s(1)$ from the two thermostats strongly depend
on $\gamma$ and both thermostats show similar dependence. However, for a given
value of $\gamma$, $D_s(1)$ from the two thermostats are not necessarily expected
to be equal and the agreement for the two thermostats for $r_c'=2.5\sigma$ is 
coincidental. As shown in Fig.~\ref{fig:tracer}, $D_s(1)$ depends not only on 
$\gamma$ but also on the cutoff $r_c'$ in Eq.~\ref{omega}. 
For small values of $\gamma$, $D_s(1)$ decreases
exponentially, while for large values of $\gamma$, $D_s(1)$ decreases as $\gamma^{-1}$
as expected since the random and dissipative forces dominate the interaction
between particles and the diffusion becomes  Brownian. Note that for large
values of $\gamma$ the Langevin thermostat becomes unstable for the value of
timestep used, $\Delta t=0.009\tau$. The strong dependence of
$D_s(1)$ on $\gamma$ is often ignored specially for DPD particles where large
values of $\gamma$ and large time steps are usually taken. In the rest of our 
simulation we choose $\gamma=0.1\tau^{-1}$ so that the interdiffusion is
only weakly affected by the random and dissipative forces.

\subsection{Concentration dependence of diffusion constants}
\begin{figure}[bth!]
\begin{center}
\includegraphics*[width=3.0in]{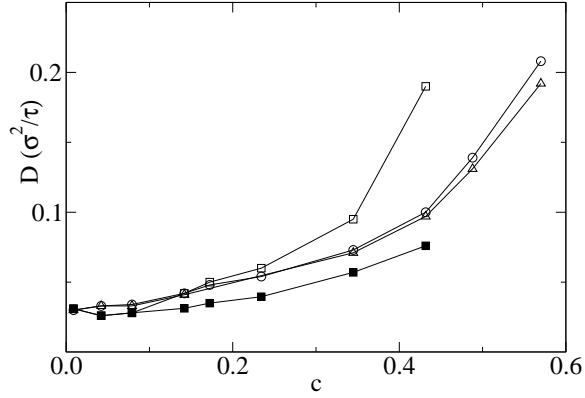}
\end{center}
 \caption{
 Dependence of diffusion constants on solvent concentration
 for $D_s(c)$ using the Langevin thermostat ($\vartriangle$) and
 DPD thermostat (o), and $D_c(c)$ ($\square$) using the DPD thermostat and
 $D(c)$ ($\blacksquare$) from Darken equation Eq.~\ref{darken}. For 
 $c\le0.09$, points for $D_c(c)$ and $D(c)$ overlap. The solid
 lines are a guide to the eye and all 
 results are for polymer chain length $N=500$ and $\epsilon_{12}=\epsilon$.
 Error bars are $\pm0.002\sigma^2/\tau$.
 }
 \label{fig:diffusion}
 \end{figure}

The self diffusion, $D_s(c)$, and corrected diffusion, $D_c(c)$, constants are calculated
from Eq.~\ref{tracer} and \ref{einstein}, respectively. The calculated values
as a function of solvent concentration are shown in Fig.~\ref{fig:diffusion}. 
For comparison, the two

thermostats with $\gamma=0.1\tau^{-1}$ were used to compute the self diffusion. 
The corrected diffusion constant was determined using the DPD thermostat. The
self diffusion values from the two thermostats are the same within the error of
the simulation. In general, for low solvent concentration both the self and 
corrected diffusion constants show weak dependence on concentration.
The self diffusion constant then increases slowly for intermediate solvent 
concentration, but a sharp increase is observed for larger concentration. 

\begin{figure}[bth!]
\begin{center}
\includegraphics*[width=3.0in]{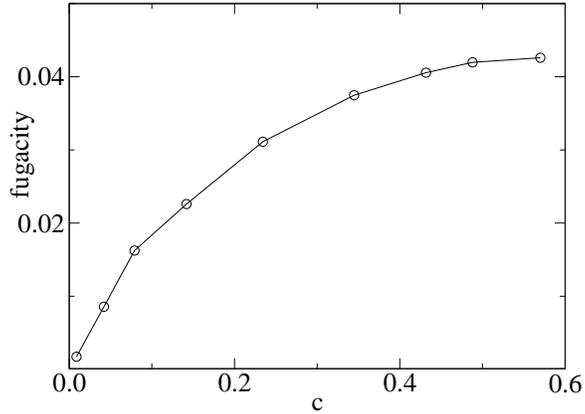}
\end{center}
 \caption{
 Fugacity as a function of solvent concentration for monomers in a polymer melt
 of chain length N=500 and $\epsilon_{12}=\epsilon$. Error bars are
 $\pm0.004$.
 }
 \label{fig:fugacity}
 \end{figure}

To calculate the diffusivity $D(c)$ from the Darken equation, Eq.~\ref{darken},
the thermodynamic factor $\partial\ln f/\partial\ln c$
is also required. The fugacity of the solvent was calculated using the GCMD
simulation method\cite{aidan} and is shown in Fig.~\ref{fig:fugacity}. 
Numerical differentiation of this data gives the required thermodynamic factor
as a function of solvent concentration. The thermodynamic factor decreases 
monotonically with concentration opposite to that obtained for
solvent in a zeolite or other porous material due to the fact that the solvent
swells the polymer, making it easier not harder to insert solvent monomer as
the solvent density increases. The diffusivity, $D(c)$, calculated from the
Darken equation is shown in Fig.~\ref{fig:diffusion}.
Note that $D(c)$ and $D_s(c)$ are equal at the dilute limit and show a similar
dependence on concentration, though $D(c)$ increases slower with increasing
concentration than $D_s(c)$.

\section{Interdiffusion}
\begin{figure}[bth!]
\begin{center}
\includegraphics*[width=4.0in]{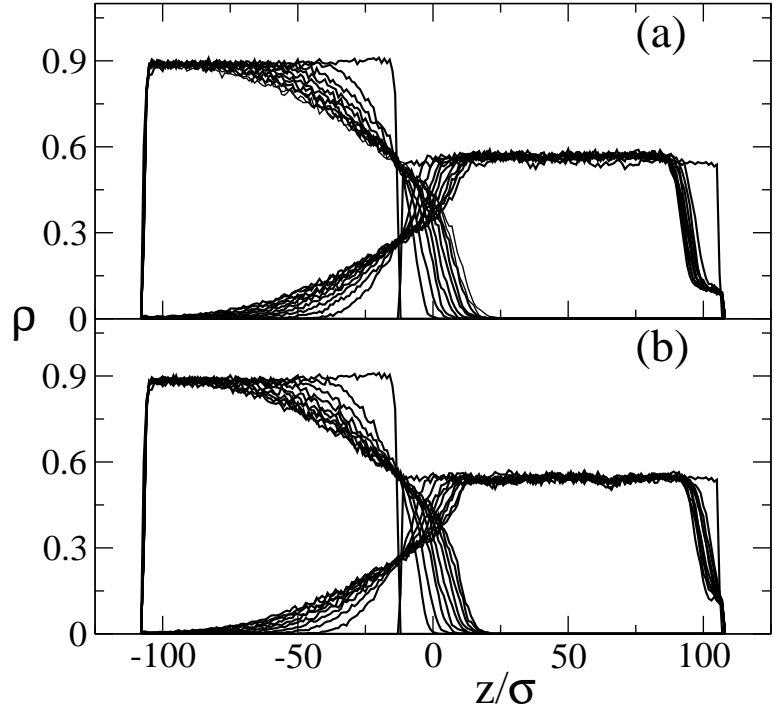}
\end{center}
 \caption{
Solvent $\rho_s$ and polymer $\rho_p$ concentration profiles as a function
of time starting from $t=0$ and plotted every 2400 $\tau$. The 
solvent is diffusing into the polymer from the right side and
(a) is for the Langevin thermostat and (b) is for the DPD thermostat.
 }
 \label{fig:interdiffusion}
 \end{figure}

\begin{figure}[bth!]
\begin{center}
\includegraphics*[width=3.0in]{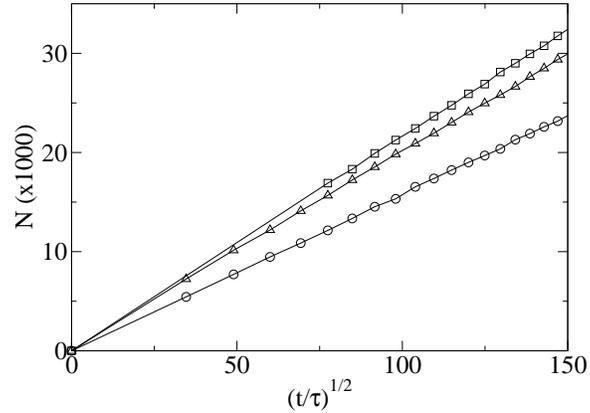}
\end{center}
 \caption{
 Weight gain for three solvent-polymer systems, monomer solvent for 
 polymer chain length N=500 ($\vartriangle$) and N=50 ($\square$) and
 dimer solvent for chain length N=500 (o) using DPD thermostat. Similar
 results for N=500 and monomer solvent were obtained for the
 Langevin thermostat.
 }
 \label{fig:weight}
 \end{figure}

\begin{figure}[bth!]
\begin{center}
\includegraphics*[width=4.0in]{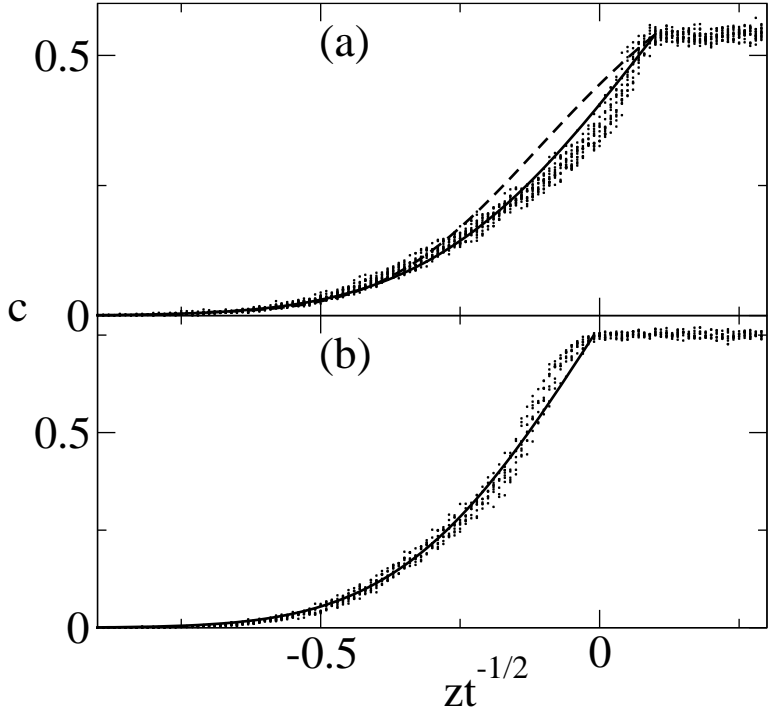}
\end{center}
 \caption{
 Solvent concentration profiles are plotted as a function
 of the scaling variable $zt^{-1/2}$, (a) for a monomer solvent that was shown
 in Fig.~\ref{fig:interdiffusion} and (b) for a dimer solvent.
The solid line in both cases is a theoretical fit
using Eq.~\ref{concentration} for the profiles. The fit using $D(c)$ from
Fig.~\ref{fig:diffusion} is shown as dashed line.
 }
 \label{fig:erf}
 \end{figure}

The initial setup for the interdiffusion studies of a solvent into an equilibrated polymer melt is 
shown in Fig.~\ref{fig:setup}. The density profile, as usually used in experiments,
of both polymer, $\rho_p$ and solvent, $\rho_s$ as a function of time for the two 
thermostats is shown in Fig.~\ref{fig:interdiffusion} for a monomer solvent
diffusing into a polymer melt of chain length $N=500$. As the solvent diffuses into the polymer, the polymer 
relaxes and the boundary is smeared out.  The density profile

at different times for the two thermostats have the same shape differing 
slightly only in time scale. The rate at 
which the solvent penetrates into the polymer can 
be determined either taking a particular value of $\rho_s$
to define the depth of penetration or by the weight gain by the 
polymer system as a function of time.
The weight gain by the polymer versus $t^{1/2}$ is shown in 
Figure~\ref{fig:weight}. Both thermostats give the same result 
confirming that the penetration increase with $t^{1/2}$ in
agreement with the Fickian diffusion. This can be further justified by
plotting the solvent density profiles of Fig.~\ref{fig:interdiffusion} for
the DPD thermostat case as
a function of $zt^{-1/2}$ as shown in Fig.~\ref{fig:erf}. 
The profiles collapse on to a single master curve confirming the
Fickian behavior. 

As a first approximation we treat $D(c)$ as a constant $D_0$ as often done
experimentally and fit the solvent
concentration profiles of Fig.~\ref{fig:erf} to the erf function given by
Eq.~\ref{concentration}. As seen from in Fig.~\ref{fig:erf}(a) (solid line), 
the erf function fit the concentration profile reasonably well particularly in 
the low solvent concentration, though overestimates slightly the rate the
monomer solvent enters the polymer film near the interface. The diffusivity extracted 
from the fit in Fig.~\ref{fig:erf}(a) is $D_0\simeq0.033\pm0.002$ $\sigma^2/\tau$ 
independent of time.  This value is within error bars in agreement with $D(c)$ in the dilute limit, 
$D(0)=0.030\pm0.002$ $\sigma^2/\tau$ from Sec. III.

Using the predicted form of $D(c)$ shown in Fig.~\ref{fig:diffusion} from the
Darken equation, Eq.~\ref{fick} can be solved numerically. 
$D(c)$ can be fit to $0.031/(1-1.15 c)$.  We find the result 
shown by the dashed line in Fig.~\ref{fig:erf}(a), which fits the simulation 
result very well at low solvent concentration and overestimates slightly more
than constant diffusivity at large concentrations. This deviation
at large concentration can be partly attributed to the swelling of the polymer
which is neglected in Eq.~\ref{fick}.
In Fig.~\ref{fig:erf}(b) we show solvent concentration profiles for the case of 
dimer solvent and due to the smaller density difference between the solvent
and the polymer melt there is less swelling. Since it is difficult to calculate
the fugacity for the dimer solvent case we did not calculate the diffusivity as
a function of concentration. In this case a constant diffusivity fit (solid 
line) gives very good agreement with the simulation results even at large 
concentration. The diffusivity extracted from this fit is 
$D_0=0.017\pm0.001$ $\sigma^2/\tau$, which also equaled
$D_s(0)$. Note that the increase in $D_s(c)$ from the dilute to the pure solvent
limit is about a factor of 6 for the monomer solvent case and a factor of
3.5 for the dimer solvent case. An extrapolation of the fit to the monomer
$D(c)$ gives a factor of 2.9 from the dilute to the pure solvent limit
which suggest that for the dimer case $D(c)$ should increase very slowly with
concentration, which further supports the assumption of constant diffuivity.

Our result agrees with the recent hypothesis,\cite{kuntz} based on two-dimensional lattice gas
automation simulation, that the nature of the diffusion is not related to the
solvent concentration gradient within the system but to the diffusivity
gradient ($dD(c)/dc$), and the standard Fickian diffusion only occurs when
$dD(c)/dc\approx0$. This justifies the experimental fits to the erf function.

We also explored interdiffusion in a number of different ways, including
varying $\epsilon_{12}$ (polymer-solvent interaction parameter) and polymer
chain length. First consider the effect of varying 
$\epsilon_{12}$. For $\epsilon_{12}=0.8\epsilon$ the solvent did not diffuse
into the polymer melt for the time scale of our simulation. In this case the
solvent can be considered to be a poor solvent. On the other hand, for 
$\epsilon_{12}=1.2\epsilon$
little change from the previous Fickian diffusion behavior was observed.
For this case the diffusion constant extracted from the error function fit is
slightly larger, $D_0=0.037\pm0.003$ $\sigma^2/\tau$,than for $\epsilon_{12}=1.0$.

To study the effect of the chain length of the polymer we decreased the chain
length from $N=500$ to 50. As seen from Fig.~\ref{fig:weight} the diffusion
process did not change as expected.  The diffusion constant $D_0$ for a monomer
solvent extracted
from the error function fit for $N=$50 is $D_0=0.035\pm0.002$
$\sigma^2/\tau$ compared to $D_s(0)=0.032\pm0.002$ $\sigma^2/\tau$.

\section{Summary}
In this work large scale molecular dynamics and grand canonical Monte Carlo 
simulation techniques were used to study the interdiffusion of a solvent into
a polymer melt. The self and corrected diffusion constants as a function of 
solvent concentration were determined separately and compared to those obtained
from the interdiffusion studies using the Darken equation. For low and 
intermediate solvent concentration, both $D_s(c)$ and $D_c(c)$ increased slowly 
with solvent concentration and were equal within the error of the 
simulation. For larger concentrations both diffusion constants
increased rapidly with the corrected diffusion constant increasing significantly
faster than the self diffusion constant. Because the solvent swells the polymer,
the thermodynamic factor $\partial\ln f/\partial\ln c$ decreased with increased solvent
concentration resulting in a diffusivity $D(c)$ which is essentially constant
for low and intermediate concentration and increased less rapidly at high 
concentration than both $D_c(c)$ and $D_s(c)$. The observed dependence of $D(c)$ with 
concentration is opposite to that of a zeolite where $D(c)$ increases more 
rapidly with concentration than $D_c(c)$.

Fickian diffusion behavior was observed for solvent absorption into polymer
melt for all cases studied. This was verified by the $t^{1/2}$ dependence of the weight gain by the
polymer system and thus the diffusion process can be considered to be Fickian.
The concentration profile of the solvent 
fit an error function derived from Fick's second law for constant diffusivity.
The diffusivity found from this fit was found to be independent of time 
and is equal to the self diffusion constant $D_s(0)$ at the dilute limit.
Even though $D(c)$ is not constant over the entire range of concentration,
since it varied little in the low concentration region relevant for
interdiffusion, assuming $D(c)$ constant is a very good approximation.

We also studied the dependence of the interdiffusion on the polymer-solvent
interaction strength, the chain length of the polymer and the chain length of
the solvent. When the interaction parameter was slightly lowered from the neutral
case the solvent did not diffuse into the polymer on the time scale of our 
simulation. On the other hand, increasing the polymer-solvent interaction 
parameter by the same
amount did not considerably affect the diffusion process. The diffusion process
was not also affected by a change in the chain length of the polymer. 

Future work will study the crossover from Fickian to non-Fickian diffusion as
the state of the polymer changes from a melt to a glass.

\section{Acknowledgments}
We are thankful to A. P. Thompson for providing the GCMD code and for helpful 
discussions. Sandia is a multiprogram laboratory operated by Sandia Corporation,
a Lockheed Martin company, for the United States Department of Energy's National
Nuclear Security Administration under Contract No. DE-AC04-94AL85000.









\clearpage

\clearpage
\end{document}